\documentclass[preprint]{aastex}
\usepackage{epsf}

\slugcomment{To be published in ApJ, March 20, 2000 (Vol.532)}

\def\be{\begin{equation}}
\def\ee{\end{equation}}
\def\ba{\begin{eqnarray}}
\def\ea{\end{eqnarray}}
\def\go{\mathrel{\raise.3ex\hbox{$>$}\mkern-14mu
             \lower0.6ex\hbox{$\sim$}}}
\def\lo{\mathrel{\raise.3ex\hbox{$<$}\mkern-14mu
             \lower0.6ex\hbox{$\sim$}}}
\def\bfe{{\bf e}}
\def\bu{{\bf u}}
\def\bx{{\bf x}}
\def\bO{{\bf \Omega}}

\begin{document}

\title{Tidal Interaction between a Fluid Star and a Kerr Black Hole
in Circular Orbit}

\author{Paul Wiggins\footnote{Also Department of Applied and Engineering
 Physics, Cornell University.}~  and Dong Lai}
\affil{Center for Radiophysics and Space Research, Department of
Astronomy, Cornell University,
Ithaca, NY 14853\\
E-mail: wiggins,dong@spacenet.tn.cornell.edu}

\begin{abstract}
We present a semi-analytic study of the equilibrium models of
close binary systems containing a Newtonian fluid star (mass $m$ and
radius $R_0$) and a Kerr black hole (mass $M$) in circular orbit. 
We consider the limit $M\gg m$ where spacetime is described by the Kerr 
metric. The tidally deformed star is approximated by an 
ellipsoid, and satisfies the polytropic equation of state.
The models also include fluid motion in the stellar interior,
allowing binary models with nonsynchronized stellar spin
(as expected for coalescing neutron star--black hole binaries)
to be constructed. Our relativistic, compressible Roche-Riemann model 
is a generalization of the incompressible, corotating Roche model 
studied earlier by Fishbone (1973). 

Tidal disruption occurs at orbital radius 
$r_{\rm tide}\sim R_0(M/m)^{1/3}$, but the dimensionless
ratio $\hat r_{\rm tide}=r_{\rm tide}/[R_0(M/m)^{1/3}]$ depends 
on the spin parameter of the black hole as well as on the equation of state
and the internal rotation of the star. We find that the general 
relativistic tidal field disrupts the star at a larger $\hat r_{\rm tide}$ 
than the Newtonian tide; the difference is particularly prominent if the 
disruption occurs in the vicinity of the black hole's horizon.
In general, $\hat r_{\rm tide}$ is smaller for a (prograde
rotating) Kerr black hole than for a Schwarzschild black hole. 

We apply our results to coalescing black hole--neutron star
and black hole--white dwarf binaries.
The tidal disruption limit is important for characterizing the expected
gravitational wave signals and is relevant for determining the energetics of
gamma ray bursts which may result from such disruption.

\end{abstract}

\keywords{stars: neutron -- stars: black hole -- stars: white dwarf --
gravitation -- hydrodynamics -- gamma ray burst}

\section{Introduction}

The problem of tidal disruption of fluid stars by black holes in close
binaries was first studied in the 1970s (e.g., Fishbone 1973; Mashhoon 1975).
Lattimer \& Schramm (1976) specifically considered neutron star disruption,
and focused on mass ejection in such an event and its implications for
nucleosynthesis. In recent years, coalescing binary systems containing a black
hole (BH) and a neutron star (NS) or a white dwarf (WD) have become
increasingly important both because they are one of the most promising sources 
of gravitational waves (GW) to be detected by ground-based
laser interferometric GW detectors (LIGO/VIRGO/TAMA/GEO)
or spaceborne interferometers such as LISA
(see Thorne 1997 for a review), and because they potentially serve as a
central engine for gamma ray bursts (e.g., Narayan, Paczynski \& Piran 1992;
Meszaros, Rees \& Wijers 1998; Piran 1998).

Current numerical simulations of NS-BH binaries and WD-BH binaries
(Kluzniak \& Lee 1998; Fryer et al.~1998; Uryu \& Eriguchi 1999;
Janka et al.~1999) are based on Newtonian theory (in some cases, the
gravitational field of the BH is approximated by a pseudo-Newtonian 
potential), and mostly (with the exception of Janka et al.~1999)
neglect gravitational radiation reaction on the dynamics. 
While such simulations provide useful insight, full understanding of 
the coalescence will have to wait for the development of numerical
relativity.

In this paper, we develop a semi-analytic model of close binaries 
containing a fluid star (mass $m$) and a massive BH (mass $M$).
To make the problem tractable, we make a series of approximations.
The obvious simplification is that in the limit $M\gg m$, 
the spacetime can be approximated by the Kerr metric. 
Furthermore, we assume that the radius of the companion star, $R_0$, is 
much greater than $Gm/c^2$, allowing the self gravity to be treated
in Newtonian theory. The companion star is modeled 
using the standard ellipsoid ansatz (Chandrasekhar 1969) and its compressible
generalization (Lai, Rasio \& Shapiro 1993,1994;
hereafter LRS1 and LRS2). This has the advantage that it reduces the
infinite degrees freedom and the partial differential equations to a set of
coupled ordinary differential equations. These equations are then reduced to
coupled algebraic equations by demanding quasi-equilibrium. Fishbone (1973) has
constructed ``relativistic Roche models'' where the incompressible fluid body 
is tidally locked with the orbital motion. However, it is necessary
to consider models where internal fluid motion is allowed, since
in a coalescing binary the NS spin is not expected to be synchronized
with the orbit (Kochanek 1992; Bildsten \& Cutler 1992). 
Newtonian Roche-Riemann binary model has been studied in LRS1.
The ``relativistic Roche-Riemann model'' (the subject of
this paper) represents a generalization to these previous works
by allowing for nonsynchronized stellar spin and including general 
relativistic tidal field. We note that Shibata (1996) has studied some aspects
of the relativistic Roche-Riemann model where the fluid body is assumed
incompressible. Post-Newtonian correction in binary ellipsoid models 
has also been studied (e.g., Taniguchi \& Shibata 1997; 
Shibata \& Taniguchi 1997; Taniguchi 1999). 

Our relativistic Roche-Riemann models (and other semi-analytic
models based on ellipsoid) clearly have limitations (see also \S 7).
For example, at small orbital radii, higher-order tidal fields are expected
to be important and the star will not have an exact ellipsoid shape.
In the case of NS-BH binaries, the conditions $M/m\gg 1$ and
$R_0c^2/(Gm)\gg 1$ are satisfied only approximately. However, given that fully
relativistic numerical calculations are still prohibitively difficult, our
analytic models have the virtue in that they provide a wide array of
semi-quantitative information about the behavior of the these systems. For
example, some of the
questions we address in this paper are: How does the tidal 
limit in general relativity compare with that in Newtonian gravity? 
How does the tidal limit depend on the spin parameter of the BH (and other 
parameters related to the NS equation of state and internal rotation)?
The answers to these questions are important for determining the
gravitational wave signals and the energetics of the gamma ray bursts
that may result from the tidal disruption. While the absolute value of
the tidal limit calculated from our model may not be accurate, its
dependence on various parameters should be robust (see \S 7), and can be 
used to calibrate results from numerical simulations when they become
available. 

In \S 2 and \S 3 we present the basic equations of our model and derive
the scaling relations. Numerical results are given in \S 4, where we
focus on the tidal limits for different models. The tidal effect 
on the orbital dynamics is considered in \S 5, and in \S 6 we
discuss possible applications of our results to gamma ray bursts.
In \S 7 we comment on the validity and limitations of our models. 
Throughout the paper we shall use units such that $G=c=1$.

\section{Basic Equations}

\subsection{Tidal tensor}

\begin{figure}[hbt]
\begin{center}
\leavevmode
\hbox{
\epsfysize=5.0in
\epsffile{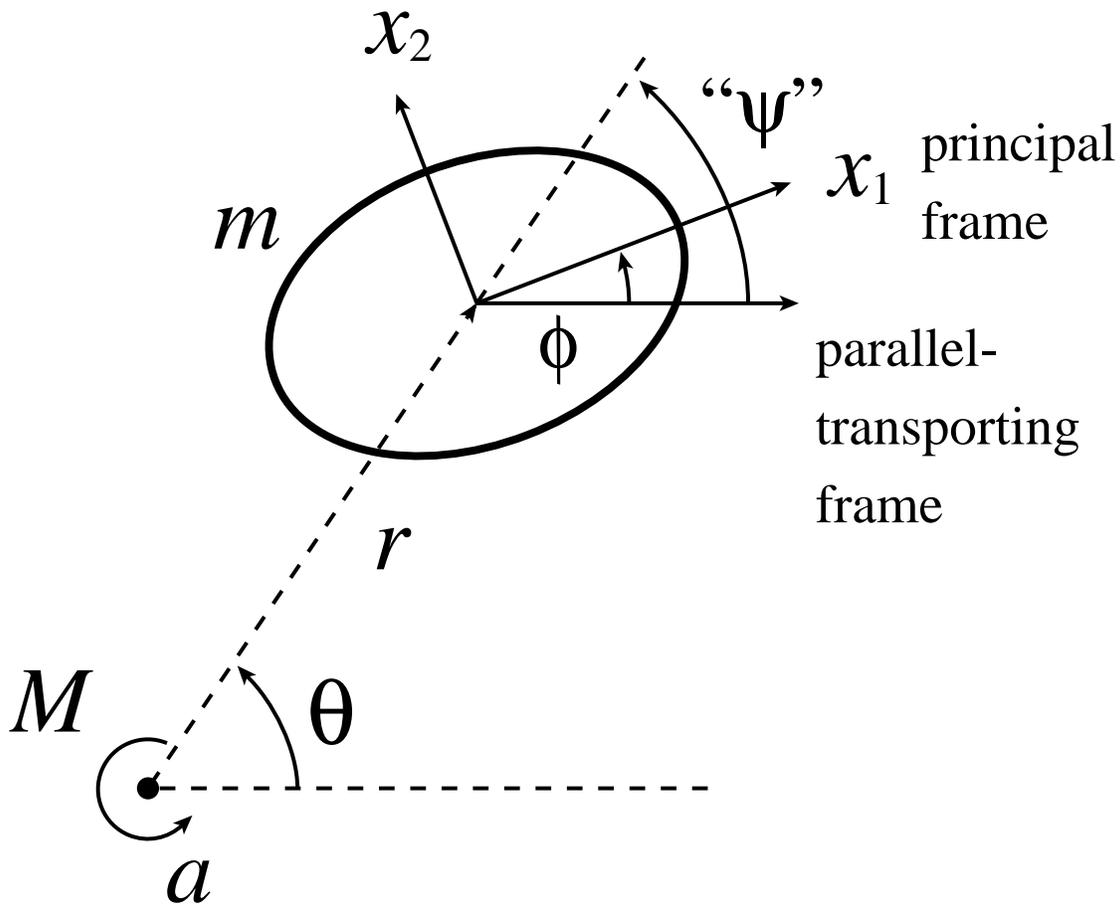}}
\caption{
Sketch of the Roche-Riemann binary model. The star is represented by 
an ellipsoid with mass $m$. The BH has mass $M$ and Kerr parameter
$a$ (The BH spin is aligned with the orbital angular momentum). 
The angle $\Psi$ is shown with quotes since the drawing is only 
correct in the Newtonian limit (see text).   
\label{fig:varfig}}
\end{center}
\end{figure}

The formalism of calculating tidal tensor in general relativity
has been studied by many authors (e.g., Fishbone 1973; Mashhoon 1975;
Marck 1983). The basic calculations and results are briefly
summarized as follows. Consider a body (with mass $m$ much smaller than the BH
mass $M$) on a geodesic in the Kerr metric.
(i) One starts with the equation of geodesic deviation:
\be
{d^2n^\alpha\over d\tau^2}+ R^\alpha_{\beta\gamma\delta}
u^\beta n^\gamma u^\delta=0,
\ee
where $\bu$ is the 4-velocity of the center-of-mass, ${\bf n}$ is the 
4-separation vector, $R^\alpha_{\beta\gamma\delta}$ 
is the Riemann tensor, which can be calculated in
certain convenient basis;
(ii) Construct a parallel-transporting orthonormal
basis carried by the body: $\{\bfe_{(0)},\bfe_{(1)},\bfe_{(2)},\bfe_{(3)}\}$,
with $\bfe_{(0)}=\bu$, and satisfies the parallel transport equation
$\nabla_{\bf u}\bfe_{(a)}=0$ ($a=0,1,2,3$);
(iii) The geodetic deviation equation can be reduced to
\be 
{d^2X_i\over d\tau^2}+\sum_j C_{ij}X_j=0,~~~~~(i=1,2,3),
\ee
where $X_i\equiv \bfe_{(i)\mu}n^\mu$, and the tidal tensor is
\be
C_{ij}=R_{\alpha\beta\gamma\delta}\bfe_{(0)}^\alpha
\bfe_{(i)}^\beta \bfe_{(0)}^\gamma \bfe_{(j)}^\delta.
\ee
We set up the parallel-transporting frame as in Figure~\ref{fig:varfig}, with 
$\bfe_{(3)}$ perpendicular to the orbital plane.
The nonzero components of the tidal tensor in this frame is given by (Marck
1983)
\begin{eqnarray}
C_{11}&=&{M\over r^3}\left(1-3{r^2+K\over r^2}\cos^2 \Psi\right),\label{c11}\\
C_{22}&=&{M\over r^3}\left(1-3{r^2+K\over r^2}\sin^2 \Psi\right),\\
C_{33}&=&{M\over r^3}\left(1+3{{K}\over {r^2}} \right), \\
C_{12}=C_{21}&=&{M\over r^3}\left(-{3\over 2}{{r^2+K}\over r^2} 
\sin{2\Psi} \right).\label{c12}
\end{eqnarray}
Here the angle $\Psi$ is specified by the equation 
\be
\dot\Psi={d\Psi\over d\tau}={E(L_z-aE)+a\over r^2+K},
\ee
where $E$ is the the energy (per unit mass), $L_z$ is the $z$-orbital 
angular momentum (per unit mass), $K=(L_z-aE)^2$ is the Carter constant (with
$Q=0$), and the dot (``$\cdot$'') denotes 
$d/d\tau$ in this and all the following equations.  
In circular orbit, we have (e.g., Misner, Thorne, \& Wheeler 1973)
\ba
E&=&{r^2-2Mr+a\sqrt{Mr}\over r\sqrt{P}},
\label{equation:energy}\\
L_z&=&{\sqrt{Mr}(r^2-2a\sqrt{Mr}+a^2)\over r\sqrt{P}},
\ea
where $P=r^2-3Mr+2a\sqrt{Mr}$. Thus we have
\be 
K=(aE-L_z)^2,\qquad\quad
\dot\Psi=\left({M\over r^3}\right)^{1/2}.
\label{equation:psiorbit}
\ee
Note that in general relativity, the angle $\Psi$ is not equal to
$\theta$, the true anomaly of the orbit, which satisfies
\be
\dot \theta = \left[{2M \over r}aE+\left(
1-{2M\over r}\right)L_z \right]\left(r^2+a^2-2Mr\right)^{-1}.
\ee

\subsection{Fluid Equations}

Including internal pressure and self-gravity, the Euler equation
in the parallel-transporting frame reads
\be
{d V_i\over d\tau}=-{1\over\rho}{\partial P\over\partial X_i}
-{\partial\Phi\over\partial X_i}-\sum_j C_{ij}X_j,
\ee
where $V_i=dX_i/d\tau$ and $\Phi$ is the potential for the (Newtonian)
self-gravity. It is convenient to set up another frame, tied to the principal
axis of the body (see Fig.~\ref{fig:varfig}). This ``principal frame'' 
is related to the parallel-transporting frame by a rotation of
angle $\phi$:
\be
x_i=\sum_j T_{ij}X_j,
\ee
where $T_{11}=T_{22}=\cos\phi$, $T_{12}=-T_{21}=\sin\phi$,
$T_{33}=1$, and the other elements are zero.
In the principal frame, the Euler equation becomes:
\be
\ddot\bx+2\bO\times\dot\bx+\bO\times (\bO\times\bx)+\dot\bO\times\bx
=-{1\over\rho}\nabla P-\nabla\Phi-\nabla \Phi_{\rm tide},
\label{euler}\ee
where
\be
\label{equation:quad_term}
\Phi_{\rm tide}={1\over 2}\sum_{ij}c_{ij}x_ix_j.
\ee
The tidal tensor in the principal frame is
\be
c_{ij}=\sum_{kl}T_{ik}T_{jl}C_{kl}.
\ee
The nonzero components are:
\begin{eqnarray}
c_{11}&=&{M\over r^3}\left[1-3{r^2+K\over r^2}\cos^2 (\Psi-\phi)\right],
\label{equation:tidal_terms_first}\\
c_{22}&=&{M\over r^3}\left[1-3{r^2+K\over r^2}\sin^2 (\Psi-\phi)\right],\\
c_{33}&=&{M\over r^3}\left(1+3{{K}\over {r^2}} \right), \\
c_{12}=c_{21}&=&{M\over r^3}\left[-{3\over 2}{{r^2+K}\over r^2} 
\sin{2(\Psi-\phi)} \right].
\label{equation:tidal_terms_last}
\end{eqnarray}

\subsection{Fluid Equations in the Ellipsoid Representation}

Since the tidal force is a linear function of $x_i$, for an incompressible 
fluid, ellipsoidal figures describe exactly the solution of the fluid
equation (\ref{euler}) (Chandrasekhar 1969). We can further generalize 
the ellipsoid figures to compressible fluid by assuming 
that the surfaces of constant density inside the star form self-similar
ellipsoids and that the the velocity of a fluid element is a linear 
function of $x_i$ (LRS1,LRS2). These approximations reduce the 
infinite degrees of freedom and partial differential equations, 
governing the dynamics of the companion, to five
dynamic variables associated with the body and a set of ordinary 
differential equations.  The five dynamic variables are the three principal
axes $a_1$, $a_2$, $a_3$ and two angles $\phi$, $\lambda$, defined such 
that $d\phi/d\tau=\Omega$ and $d\lambda/d\tau=\Lambda$.
Here $\Omega$ is the angular velocity of the ellipsoidal figure measured 
in the parallel transporting frame, and $\Lambda$ characterizes the 
internal fluid motion (see eq.~[\ref{vfield}] below) via
$\Lambda=-a_1a_2\zeta/(a_1^2+a_2^2)$ with $\zeta$ the (uniform) vorticity
along the $z$-axis in the frame corotating with the figure. 
The physical significance of $\phi$ is illustrated in Fig.~\ref{fig:varfig}, 
but $\lambda$ is without meaning and only its derivative, $\Lambda$, appears in
our equations. The orbit is specified by the orbital separation $r$ and 
the true anomaly $\theta$. 

Several parameters must be given to complete the description of the system:
In addition to the stellar mass $m$, the BH mass $M$ (satisfying 
$M\gg m$) and the Kerr parameter $a$, there are parameters associated 
with the equation of state and the initial stellar rotation. 
We adopt the polytropic equation of state,
\be
P=K\rho^{1+1/n},
\ee
with $n$ the polytropic index. Note that $K$ 
(not to be confused with that $K$ in the tidal tensor; see
eqs.~[\ref{c11}]-[\ref{c12}]) is uniquely determined from 
$n,~m$ and $R_0$, the radius of the nonrotating star with the same mass.
Thus instead of $K$, we shall use $R_0$ as a parameter. Finally,
we allow the star to have some initial rotation, 
$\Omega_s$, when the binary separation is infinite (see \S 4.3).
  
The equations governing the dynamics of the system can be obtained
using the variational principle. It is convenient to write the Lagrangian 
as the sum of two terms:
\be
L = L_{\rm body} + L_{\rm tide}.
\ee    
We assume that the star moves on a circular geodesic around the BH, 
and neglect tidal effect on the orbital motion --- This is an excellent
approximation for $M\gg m$ even if the star is highly distorted (see \S 5).
Thus the orbital terms do not appear in the Lagrangian.  
The internal Lagrangian of the body, $L_{\rm body}$, is exactly the same as 
in the Newtonian treatment. The velocity field ${\bf u}$
of the fluid inside the star relative to its center of mass can be written as
\be
{\bf u} = \left[\left({a_1\over a_2}\Lambda-\Omega\right)x_2{\bf e}_1 +
\left(-{a_2\over a_1}\Lambda+\Omega\right)x_1{\bf e}_2\right]+
\left({{\dot a_1}\over a_1}x_1{\bf e}_1+{{\dot a_2}\over a_2}x_2{\bf e}_2+
{{\dot a_3}\over a_3}x_3{\bf e}_3\right),
\label{vfield}\ee
where $\bfe_1,~\bfe_2$ and $\bfe_3$ are the basis unit vectors along the
principal axes of the ellipsoid. 
On integration, the kinetic energy of the star relative to its center 
of mass is given by
\be
T={1\over 2}I(\Lambda^2+\Omega^2)-{2\over 5}\kappa_n m a_1 a_2 \Lambda
\Omega+{1\over 10}\kappa_n m (\dot a_1^2+\dot a_2^2+\dot a_3^2),
\ee
with $I=\kappa_n m (a_1^2+a_2^2)/5$ and where $\kappa_n$ is a constant
depending on $n$ and specifies the mass concentration inside the star
(see Table 1 in LRS1).  The internal energy is:
\be
U=\int\!{dm\, n {P \over \rho }}=k_1 K\rho_c^{1/n} m
\ee
where $k_1$ is another constant depending only on $n$, and $\rho_c
\propto m/(a_1a_2a_3)$ is the central density.  The self gravitational
potential energy is given by:
\be
W=-{3\over{5-n}}{m^2\over R}{{\cal I}\over{2R^2}},\ {\rm with}\ 
{\cal I}=A_1a_1^2+A_2a_2^2+A_3a_3^2,
\ee   
where $R \equiv (a_1a_2a_3)^{1/3}$ is the mean radius of the ellipsoid,
and the dimensionless index symbols $A_i$ are defined as in Chandrasekhar
(1969; \S 17). The Lagrangian for the star is then
\be
L_{\rm body} = T-U-W.
\ee
Using the tidal tensor derived earlier (see eq.~[\ref{equation:quad_term}]),
the tidal Lagrangian may be calculated as
\be
L_{\rm tide}=-W_{\rm tide}=-\int\!d^3x\,\rho(\bx)\Phi_{\rm tide}(\bx)
=-{1\over 2}(I_{11} c_{11}+I_{22} c_{22}+ I_{33} c_{33}),
\label{wtide}\ee
with 
$I_{ij}={{\kappa_n m a_i^2 \delta_{ij}}/5}$.

The dynamic equations of the system are found using the Euler-Lagrange 
equation:
\be
{d\over {d\tau}} \left( {{\partial L}\over{\partial \dot q_i}} \right) =
{{\partial L}\over{\partial q_i}},  
\ee
where $q_i$ represents one of the dynamic variables 
$\{a_1,a_2,a_3,\phi,\lambda\}$. We find
\begin{eqnarray}
\label{equation:alg1}
\ddot a_1&=&a_1(\Omega^2+\Lambda^2)-2a_2\Omega \Lambda -{{2 \pi} \over {q_n}}
a_1 A_1 \overline{\rho} +{{5k_1 P_c}\over{n\kappa_n \rho_c a_1}}-c_{11} 
a_1, \\
\label{equation:alg2}
\ddot a_2&=&a_2(\Omega^2+\Lambda^2)-2a_1\Omega \Lambda -{{2 \pi} \over {q_n}}
a_2 A_2 \overline{\rho} +{{5k_1 P_c}\over{n\kappa_n \rho_c a_2}}-c_{22} 
a_2, \\
\ddot a_3&=&-{{2 \pi} \over {q_n}}
a_3 A_3 \overline{\rho} +{{5k_1 P_c}\over{n\kappa_n \rho_c a_3}}-c_{33} 
a_3,\label{equation:alg3}\\
\label{equation:ang_mo}
\dot J_s &=& {3\over 10}\kappa_n m {M\over r^3} {{r^2+K}\over r^2}
(a_1^2-a_2^2)\sin\left[2(\Psi-\phi)\right],\label{dotjs}\\
\dot {\cal C} &=& 0,
\end{eqnarray}
where we have defined
\ba
J_s &\equiv& I\Omega-{2 \over 5 } \kappa_n m a_1 a_2 \Lambda, \\
{\cal C} &\equiv& I\Lambda-{2 \over 5 } \kappa_n m a_1 a_2 \Omega,
\ea
and $\overline{\rho} \equiv 3m/(4\pi R^3)$, $q_n \equiv \kappa_n (1-n/5)$.
The quantity $J_s$ is the spin angular momentum of the 
star and ${\cal C}$ is interpreted as the circulation in the locally
nonrotating inertial frame. In the absence of viscosity,    
Kelvin's theorem ensures that the circulation is conserved. 
Note that the corotation case corresponds to $\Lambda=0$. 

For a binary system in quasi-equilibrium, we require
\be
\ddot a_i=0,\ \phi=\Psi,\ {\rm and } \ 
\dot\phi=\Omega=\dot \Psi=\sqrt{M\over r^3}, 
\ee
where the second equality comes from $\dot J_s=0$ and 
eq.~(\ref{dotjs}), and the third equality follows from 
(\ref{equation:psiorbit}).  
It is of interest to note that in general relativity, 
quasi-equilibrium does not require the angular frequency of the
figure, $\Omega=\dot\phi$, to be equal to the orbital angular frequency
$\dot \theta$; rather, $\Omega=(M/r^3)^{1/2}$, independent of the spin of 
the BH. 

The pressure terms in equations (\ref{equation:alg1})-(\ref{equation:alg3})
can be conveniently rewritten in terms of the dynamic variables (LRS2):
\be
\label{equation:pressure}
{{5k_1}\over{n\kappa_n}}{P_c\over \rho_c}={m\over{q_n R_0}} 
\left( {R_0\over R} \right)^{3/n}\ {\rm with}\ R=\left(
a_1a_2a_3\right)^{1/3}. 
\ee
For $n\rightarrow 0$, we have $k_1/n\rightarrow {2/5}$ and $\kappa_n
\rightarrow 1$, so that
\be
\label{equation:nzero1}
{{5k_1}\over{n\kappa_n}}{P_c\over \rho_c} \rightarrow 
{2P_c\over \rho_c}.
\ee
For an incompressible ellipsoid, volume is conserved,
$R^3=a_1a_2a_3=R_0^3$; Differentiating this with respect to $\tau$ gives
$\sum_i{{\dot a_i}/a_i}=0$.
A second derivative yields
$\sum_i{{\ddot a_i}/a_i} = \sum_i\left({{\dot a_i}/a_i}\right)^2$,
which maybe combined with eqs.~(\ref{equation:alg1})-(\ref{equation:alg3}) 
to give
\be
\label{equation:nzero2}
{2P_c\over \rho_c} = \left(\sum_i {1\over a_i^2}\right)^{-1}
\left[-2\left(\Omega^2+\Lambda^2\right)+2\Omega\Lambda\left({
a_2\over a_1}+{a_1\over a_2}\right)+4\pi\overline\rho
+ \sum_i\left({{\dot a_i}\over a_i}\right)^2\right],
\ee
where we have used the relations
$A_1+A_2+A_3=2$ and $k_{11}+k_{22}+k_{33}=0$.
Of course, for quasi-equilibrium, we have $\dot a_i=0$.

\section{Dimensionless Equations and Scaling Relations}

Before presenting our results on specific binary models,
it is useful to cast the equations into a dimensionless form. 
This will shed light on how our results depend on various parameters 
($m,~M,~a,~R_0$) of the system. 
We define the following set of dimensionless variables:
\begin{eqnarray}
\hat a_i&=&{a_i\over R_0},\\
\hat \Omega &=& \Omega \left( {m\over R_0^3} \right)^{-1/2}, \\
\hat \Lambda &=& \Lambda  \left( {m\over R_0^3} \right) ^{-1/2}, \\
\label{equation:dimlesstidal}
\hat c_{ij} &=& c_{ij}  \left( {m\over R_0^3} \right)^{-1}.  
\end{eqnarray}
Since the tidal radius is expected to be of order $R_0(M/m)^{1/3}$, 
we define the dimensionless orbital radius as
\be
\hat r \equiv {r\over R_0} \left( {m\over M} \right)^{1/3}.
\ee
In terms of these dimensionless variables, equations  
(\ref{equation:alg1})-(\ref{equation:alg3}) can be written in 
the following form (for quasi-equilibrium):
\begin{eqnarray}
\label{equation:dimless1}
0&=&\hat a_1(\hat \Omega^2+\hat \Lambda^2)-2\hat a_2\hat \Omega \hat 
\Lambda -{3\over 2} A_1 \hat a_1 \left({R_0\over R}\right)^3 +
{1\over{\hat a_1q_n}}\left({R_0\over R}\right)^{3/n}
-\hat c_{11} \hat a_1, \\
0&=&\hat a_2(\hat \Omega^2+\hat \Lambda^2)-2\hat a_1\hat \Omega \hat 
\Lambda -{3\over 2} A_2 \hat a_2 \left({R_0\over R}\right)^3 +
{1\over{\hat a_2q_n}}\left({R_0\over R}\right)^{3/n}
-\hat c_{22} \hat a_2, \\
0&=&-{3\over 2} A_3 \hat a_3 \left({R_0\over R}\right)^3 +
{1\over{\hat a_3q_n}}\left({R_0\over R}\right)^{3/n}
-\hat c_{33} \hat a_3,
\label{equation:dimless3} 
\end{eqnarray}
where $R/R_0=(\hat a_1\hat a_2\hat a_3)^{1/3}$, and
\be
\hat\Omega=\hat r^{-3/2}.
\ee
By making (\ref{equation:nzero1}) and (\ref{equation:nzero2}) 
dimensionless, the equations for the $n=0$ case may also be written
in a dimensionless form. The dimensionless circulation is
\be
\label{equation:dimless4}
\hat {\cal C}= {{\cal C}\over (m^3R_0)^{1/2}}=
{\kappa_n \over 5} \left[(\hat a_1^2+\hat a_2^2)\hat 
\Lambda-2 \hat a_1 \hat a_2 \hat \Omega\right].
\ee
In Newtonian theory, $\hat c_{ij}$ depends only on $\hat r$, thus for given
$\hat r$ and $\hat{\cal C}$, 
eqs.~(\ref{equation:dimless1})-(\ref{equation:dimless4}) completely determine
the solution. However, this is not the case in general relativity since
$\hat c_{ij}$ depends not just on $\hat r$. For example,
\be
\hat c_{11}={1\over {\hat r^3}} \left(  -2 -3{K\over r^2} \right),
\ee
(The Newtonian case corresponds to $K=0$). It is easy to see that 
relativistic effect enters through the combination $K/r^2$, which can be
expressed as
\be
{K \over r^2}={{\left[\left({r/M}\right)^{1/2}-
\hat a \right]^2} \over {\left({r/M}\right)^2-3 ({r/M}) + 2 \hat a 
\left( {r/M} \right) ^{1/2} }},
\ee
where $\hat a\equiv a/M$. Therefore, in addition to the dimensionless
Kerr parameter $\hat a$, one could introduce another ``parameter'', $r/M$,
which specifies how relativistic the orbit is. But of course this is 
not really a parameter, since $r$ varies as the binary evolves. 
A more useful parameter is 
\be
\label{eqn:newt}
\hat R_0 \equiv {R_0(M/m)^{1/3}\over M}=
{R_0\over m} \left( {m\over M} \right)^{2/3}.
\ee  
Roughly speaking, $\hat R_0$ is the ratio of the tidal radius and the 
horizon radius of the BH. This parameter is convenient because
it depends only on the mass and radius of the star and the mass of the BH.
When $\hat R_0\gg 1$, we expect that the binary will exist only
in the Newtonian regime (i.e., the star will be disrupted before it enters
into a relativistic orbit). But when $\hat R_0$ is small (of order unity), 
then the tidal disruption will be affected by general relativity. Note that 
$r/M$ can be expressed in terms of $\hat r$ and $\hat R_0$ as
\be
{r\over M}=\hat R_0\hat r. 
\ee
Thus for given $n,~ \hat r,~\hat{\cal C},~\hat a$ and $\hat R_0$, 
an equilibrium binary model is uniquely determined.

\section{Equilibrium Binary Models and Tidal Limits}

As discussed in \S 2, in the absence of viscosity, fluid circulation 
is conserved; gravitational radiation reaction, being
a potential force, also preserves the circulation.
In a coalescing binary, we can determine the 
the relative importance of viscosity by comparing the orbital
decay time $t_d=|r/\dot r|$ and the tidal synchronization time $t_{\rm syn}$
(Bildsten \& Cutler 1992). The ratio is
\be
{t_{\rm syn}\over t_d}\simeq \left({10\over\sin 2\alpha}\right)
\left({m\over R_0}\right)^{5/2}\left({M\over m}\right)^{2/3}\hat r^{1/2},
\ee
where $\alpha$ is the tidal lag angle, and is proportional to the 
viscosity. Kochanek (1992) and Bildsten \& Cutler (1992) showed that 
$t_{\rm syn}/t_d$ is greater than unity for coalescing 
NS-BH binary. Using microscopic viscosity for white dwarf matter 
(Nandkumar \& Pethick 1984 and references therein) 
we can also show that the inequality
$t_{\rm syn}/t_d\gg 1$ is well satisfied for WD-BH binaries. 
Thus for the types of systems (NS-BH and WD-BH) of interest in this paper, 
the binary will likely evolve along a quasi-equilibrium sequence with 
constant $\cal C$. We shall therefore focus on such models in this section.

\subsection{Roche-Riemann Model: Irrotational Equilibrium Sequences}

\begin{figure}[hbt]
\begin{center}
\leavevmode
\hbox{
\epsfysize=5.0in
\epsffile{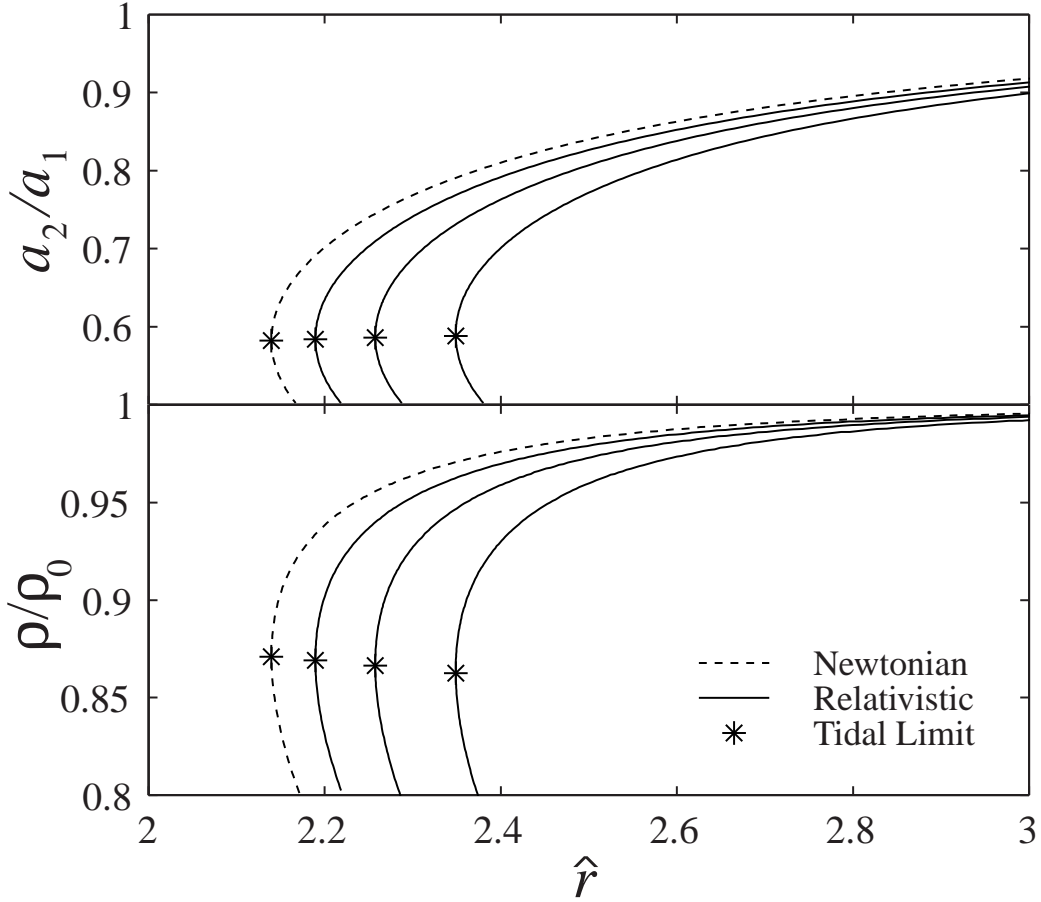}}
\caption{
Equilibrium sequences for irrotational binaries (${\cal C}=0$).
The axis ratio $a_2/a_1$ and the central density $\rho_c$
[in units of its value at infinite binary separation, $\rho_0
=\rho_c(r\rightarrow\infty)$] are plotted against 
the dimensionless orbital separation $\hat r=(r/R_0)(m/M)^{1/3}$.
The polytropic index is $n=1$. The dashed curves represent the Newtonian 
result, with $\hat R_0\rightarrow \infty$. The three solid curves include the
effects of general relativity (with $\hat a=1,\,0,\,-1$ from left to right),
all for $\hat R_0=4$. The turning point of each curve, 
with $\hat r$ reaching a minimum, is marked by an asterisk, and 
corresponds to the tidal limit. Note that the values of
$a_2/a_1$ and $\rho_c/\rho_0$ at $\hat r=\hat r_{\rm tide}$ are similar for
different sequences, but they are not identical (see Fig.~6). 
\label{fig:equil_seq}}
\end{center}
\end{figure}

For the irrotational sequence, ${\cal C}=0$, the star has zero spin
at large orbital radii ($\Omega_s=0$), and equation (\ref{equation:dimless4})
implies $\Lambda=2a_1a_2\Omega/(a_1^2+a_2^2)$. In Figure \ref{fig:equil_seq},
four equilibrium sequences are depicted illustrating the effect of general
relativity. Note that it is most convenient to parameterize the sequence
in terms of $a_2/a_1$ since $\hat r$ is double-valued.
We see that the star becomes more distorted as $\hat r$ decreases.
At the same time, its central density $\rho_c$ decreases. 
This implies that the tidal field stablizes the star against 
collapse. Physically, this stablization arises from that fact that
the self-gravity of a distorted object is weaker, and the star expands 
in order to satisfy hydrostatic equilibrium (see, e.g., Lai 1996; LRS1; 
Brady \& Hughes 1997; Wiseman 1997; Baumgarte et al.~1998; Thorne 1998;
Flanagan 1998; Bonazzola et al.~1999a,b). 

For each sequence, there exists a turning point,
corresponding to the tidal limit, $r_{\rm tide}$, below which no
equilibrium solution is possible. From Fig.~\ref{fig:equil_seq}, we see that
stars with finite $\hat R_0$'s are tidally disrupted at a larger $\hat r$ 
than purely Newtonian stars (with $\hat R_0 \rightarrow\infty$), an incarnation
of the old adage that GR means ``stronger'' gravity. We also see that the tidal
limit $\hat r_{\rm tide}$ is smaller for positive finite $\hat a$ (i.e., the
orbit is prograde with respect to the BH spin), while $\hat r_{\rm tide}$ is
larger for more negative $\hat a$. This effect of the Kerr parameter $\hat a$
might also be guessed from the effect it has on the innermost stable
circular orbit (ISCO): The ISCO is largest for $\hat a=-1$, $r_{\rm isco}=9M$,
and smallest for $\hat a=1$, $r_{\rm isco}=M$. The same qualitative effect
applies in tidal disruption; the strongest tidal gravity is seen for the $\hat
a=-1$ case.  

\begin{figure}[hbt]
\begin{center}
\leavevmode
\hbox{
\epsfysize=5.0in
\epsffile{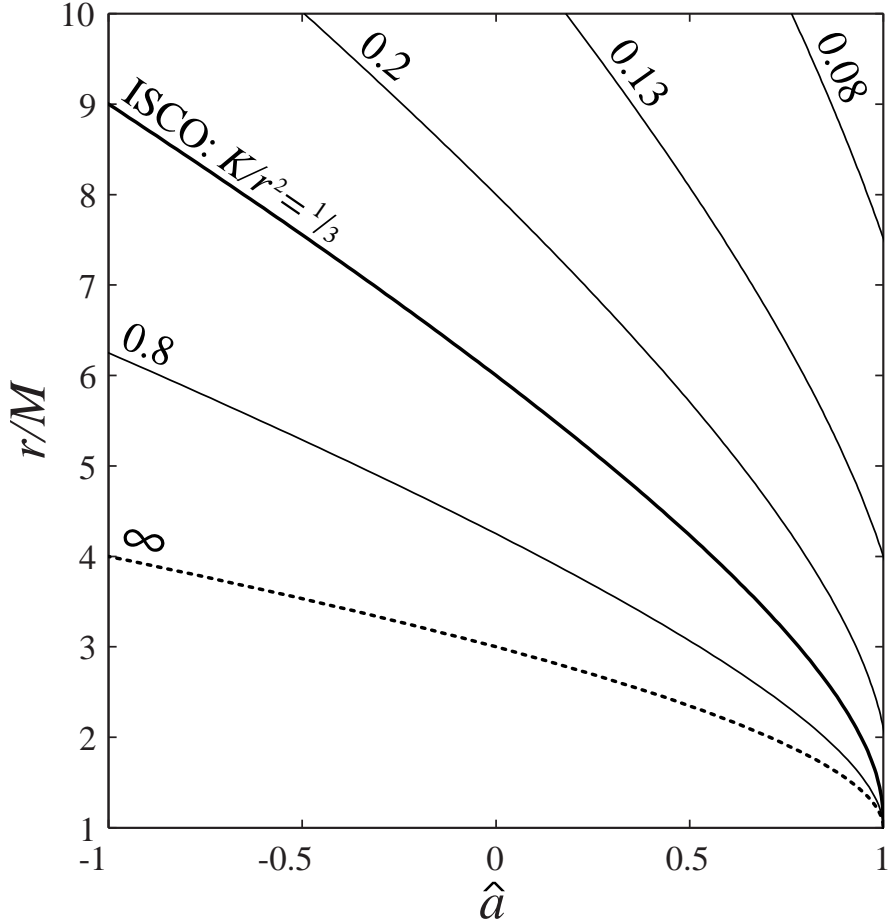}}
\caption{
Contours of constant $K/r^2$ in the $\hat a-(r/M)$ plane.
This quantity determines the magnitude 
of relativistic effect on the tidal distortion. The number 
on each curve gives the value of $K/r^2$. 
For physically reasonable $r/M$, the quantity $K/r^2$ is positive
and decreases with increasing $\hat a$ and $r/M$. 
Note that at the innermost stable circular orbit (ISCO), $K/r^2=1/3$ is
independent of $a$. Thus the heavy solid line also gives the value of
$r_{\rm isco}$ as a function of $\hat a$. The pole
of $K/r^2$ (where it becomes infinite) is hidden inside the ISCO except
when $\hat a=1$.  
\label{fig:K}}
\end{center}
\end{figure}

The dependence of the tidal limit on $\hat a$ can be understood by
examining the tidal field
(eqs.~[\ref{equation:tidal_terms_first}]-[\ref{equation:tidal_terms_last}]).
The spin parameter $\hat a$ enters the tidal field only through $K/r^2$.
Figure~\ref{fig:K} shows contours of constant $K/r^2$ in the 
$\hat a-(r/M)$ plane. Note that only the region outside the ISCO ($r>r_{\rm
isco}$) is of physical interest. At the ISCO, it can be shown that 
$K/r^2=1/3$, independent of $\hat a$ (This coincidence was first noted
by Fishbone 1973).  We see from Fig.~\ref{fig:K} that at a given $r$, 
the quantity $K/r^2$ is smaller for larger $\hat a$, implying that 
the tidal field is weaker for larger, positive $\hat a$. 

\subsection{Irrotational Tidal Limit}

As discussed above, the turning point in the equilibrium sequence is
of special significance: it corresponds to the tidal limit 
(also called ``Roche-Riemann limit''), $\hat r_{\rm tide}$.
To determine this limit, we require $\partial\hat r/\partial (a_2/a_1)=0$,
which adds another equation to be solved together with 
eqs.~(\ref{equation:dimless1})-(\ref{equation:dimless3}). 

\begin{figure}[hbt]
\begin{center}
\leavevmode
\hbox{
\epsfysize=6.0in
\epsffile{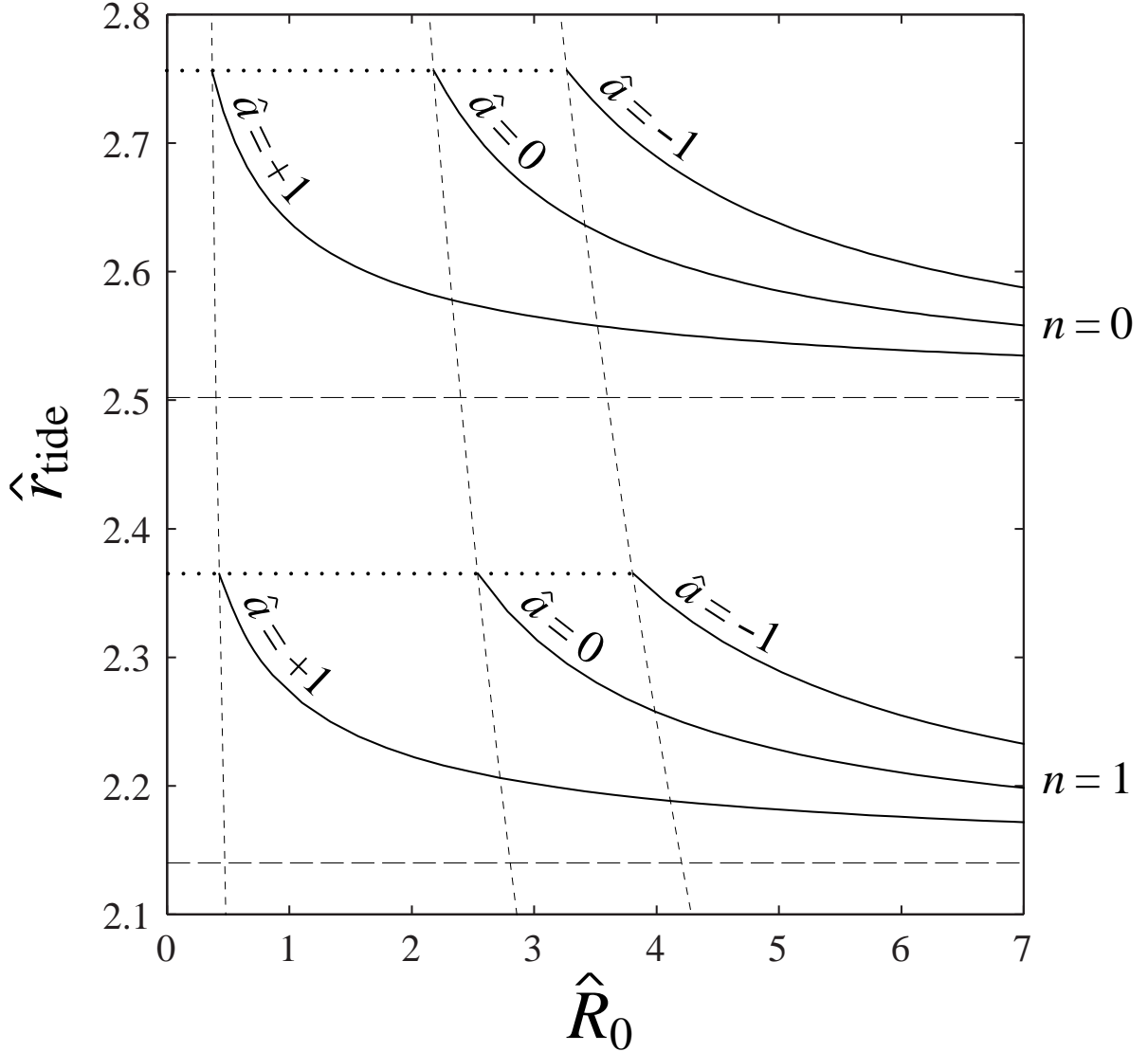}}
\caption{
Tidal limit $\hat r_{\rm tide}=(r_{\rm tide}/R_0)(m/M)^{1/3}$ 
as a function of $\hat R_0=(R_0/m)(m/M)^{2/3}$ for irrotational
binaries (${\cal C}=0$). The two sets of solid curves correspond to
polytropic index $n=0$ and $n=1$ respectively. Results for three
values of Kerr parameter ($\hat a=-1,0,1$) are shown.
Each solid curve is truncated at $\hat R_0=\hat R_{0,{\rm crit}}$,
where $r_{\rm tide}=r_{\rm isco}$
[The short-dashed lines represent the ISCO for different Kerr parameters, 
i.e., $\hat R_0\hat r=(r/M)_{\rm isco}$]. The two horizontal
dotted lines show the limiting $\hat r_{\rm tide}$ 
at the truncation. Two additional
long-dashed lines show the Newtonian limit of the tidal limit. 
For $n=0$, the Newtonian limit is $\hat r_{\rm tide}=2.502$ and the ISCO 
limit is $\hat r_{\rm tide}=2.757$. For $n=1$, the Newtonian limit is $\hat
r_{\rm tide}=2.140$ and the ISCO limit is $\hat r_{\rm tide}=2.366$. 
\label{tidal_1}}
\end{center}
\end{figure}

\begin{figure}[hbt]
\begin{center}
\leavevmode
\hbox{
\epsfysize=6.0in
\epsffile{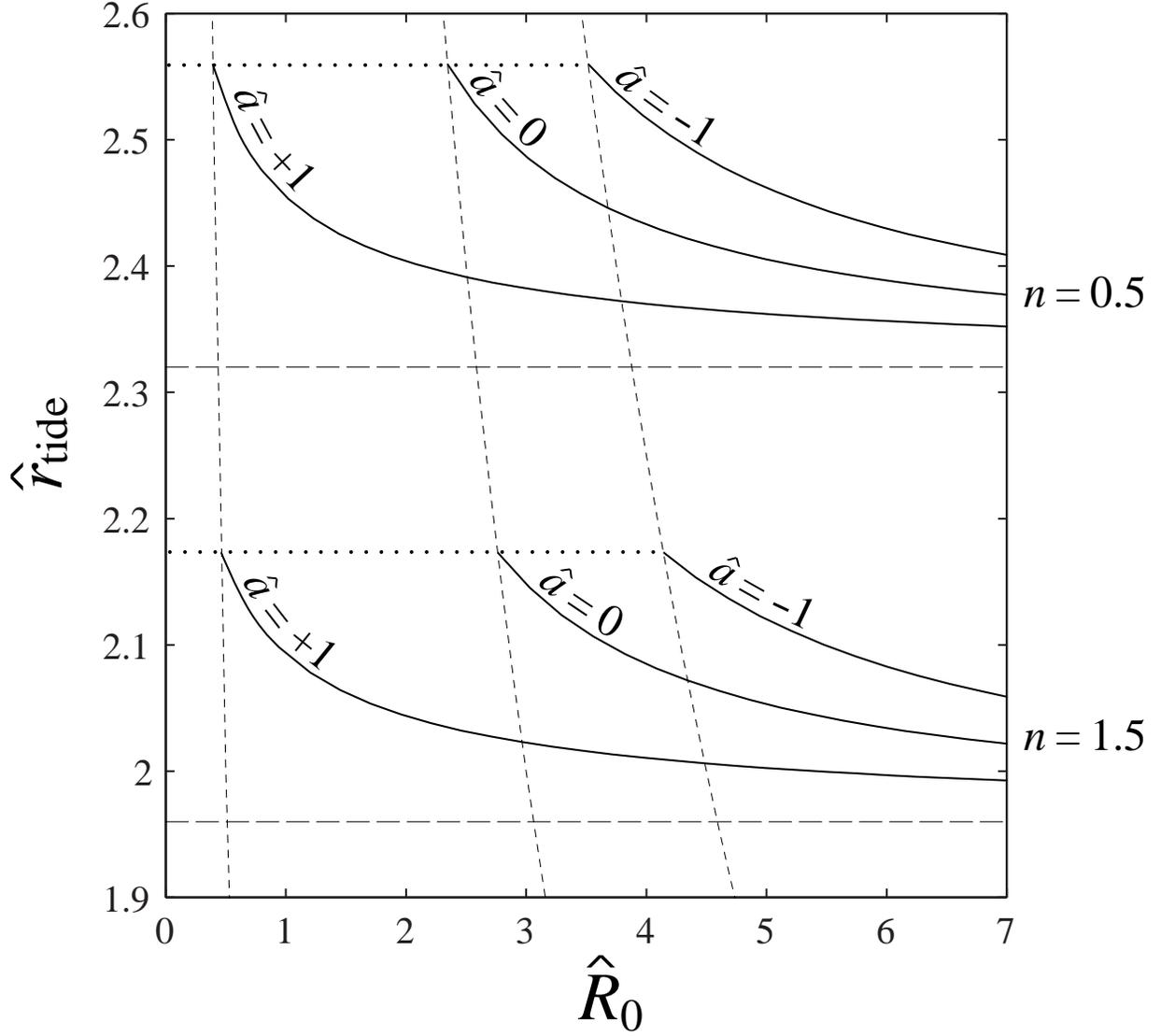}}
\caption{
Same as Fig.~4 except for $n=0.5$ and $n=1.5$.
For $n=0.5$, the Newtonian limit is
$\hat r_{\rm tide}=2.320$ and the ISCO limit is $\hat r_{\rm tide}=2.560$.
For $n=1.5$, the Newtonian limit is $\hat r_{\rm tide}=1.960$ 
and the ISCO limit is $\hat r_{\rm tide}=2.173$. 
\label{tidal_2}}
\end{center}
\end{figure}

We have calculated $\hat r_{\rm tide}$ as a function of
$\hat R_0$ for a series of polytropic indices, 
$n=\{0,0.5,1,1.5\}$ and three Kerr parameters,
$\hat a=\{-1,0,1\}$. The results are 
are shown in Fig.~\ref{tidal_1} and Fig.~\ref{tidal_2}. 
Selected numerical values are also listed in Table 1.
For given $n$ and $\hat a$, there is a critical 
$\hat R_0=\hat R_{0,{\rm crit}}$, for which
$r_{\rm isco}=r_{\rm tide}$, i.e., the inner-most stable 
circular orbit (ISCO) coincides with the tidal radius.
For $\hat R_0<\hat R_{0,{\rm crit}}$, 
the binary encounters ISCO before tidal disruption; In this case, 
tidal disruption can still occur outside the BH's event 
horizon, but it takes place on a plunging orbit rather than on 
a circular orbit as in the $\hat R_0>\hat R_{0,{\rm crit}}$ case.
Our results in Table 1 and Figs.~\ref{tidal_1}-\ref{tidal_2}
therefore have been truncated at $\hat R_0=\hat R_{0,{\rm crit}}$.
Note that for $\hat R_0=\hat R_{0,{\rm crit}}(\hat a)$ (so that tidal
disruption occurs at ISCO), the tidal limit $\hat r_{\rm tide}$ 
is independent of $\hat a$. This is because in the framework of our
calculations, general relativistic effects come into the equations only 
through $K/r^2$, which is equal to $1/3$ (independent of $\hat a$) 
when evaluated at $r_{\rm isco}(\hat a)$ (see Fig.~\ref{fig:K}). 

\begin{figure}[hbt]
\begin{center}
\leavevmode
\hbox{
\epsfysize=5.0in
\epsffile{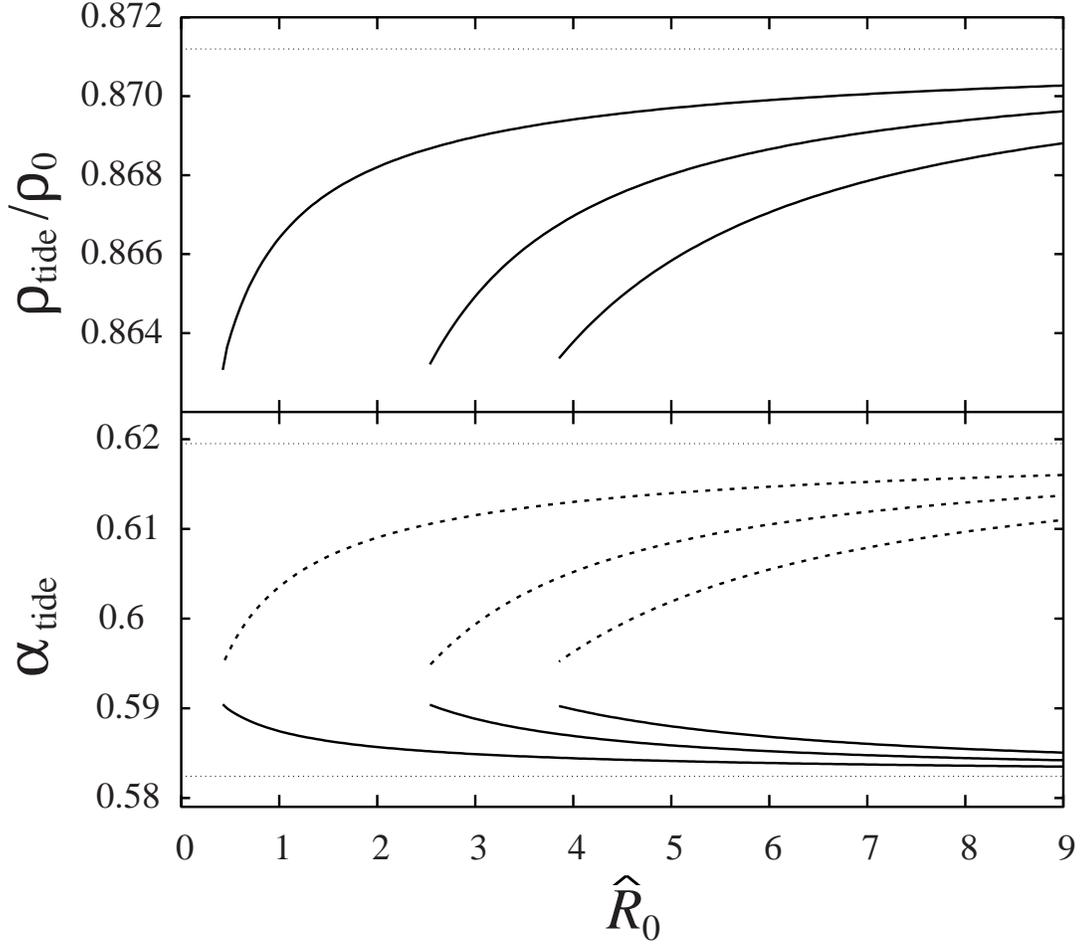}}
\caption{
The central density [in units of $\rho_0=\rho_c(r\rightarrow\infty)$]
and the axis ratio ($\alpha=a_2/a_1$ or $a_3/a_1$)
of the star at the tidal limit ($\hat r=\hat r_{\rm tide}$) as a function
of $\hat R_0$ for $\hat a=\{1,0,-1\}$, $n=1$ and ${\cal C}=0$.
The solid curves in the lower plot are for $\alpha \equiv a_2/a_1$ and the
dashed curves are for $\alpha \equiv a_3/a_1$. 
From left to right, the curves are $\hat a = \{1,0,-1\}$. 
The horizontal dotted lines correspond to the Newtonian limit.
\label{fig:distortion}}
\end{center}
\end{figure}

Figure \ref{fig:distortion} depicts the central density $\rho_c$
and the distortion ($\alpha=a_2/a_1$ or $a_3/a_1$) of the star at the tidal
limit as a function of $\hat R_0$. Only the $n=1$ results are shown as an
example. We see that $\rho_{\rm c,tide}$ and $\alpha_{\rm tide}$ (evaluated at
$r=r_{\rm tide}$) are rather insensitive to $\hat R_0$ and $\hat a$, although 
the values of $\hat r_{\rm tide}$ for different $\hat R_0$ and
$\hat a$ can differ by as much as $10\%$ (see Table 1 and 
Figs.~\ref{tidal_1}-\ref{tidal_2}).

\subsection{The Effect of Initial Spin}

When the star has an intrinsic spin $\Omega_s$ at large orbital radii, 
its (conserved) circulation is finite. Since at $r\rightarrow\infty$,
the star is axisymmetric ($a_1=a_2$), and the figure rotation
$\Omega\rightarrow 0$, we have (see eq.~[\ref{equation:dimless4}])
\be 
\hat {\cal C} = -{2\over 5} \kappa_n \hat a_1^2 \hat \Omega_s,
\ee     
where we have identified $\Omega_s=-\Lambda(r\rightarrow\infty)$. 
Note that when $\Omega_s$ is positive (i.e., the stellar spin is 
in the same direction as the orbital angular momentum), $\cal C$ is 
negative. The maximum spin that a uniformly rotating star can sustain without 
shedding mass from its equator is given by
(e.g., Friedman, Ipser \& Parker 1986; Cook, Shapiro \& Teukolsky 1992)
\be
\hat\Omega_s={\Omega_s\over (m/R_0^3)^{1/2}}\lo 0.6.
\ee
For a given $\hat \Omega_s$, the dimensionless circulation ${\cal C}$
can be calculated by using the equations for compressible Maclaurin 
spheroid (LRS1). 

\begin{figure}[hbt]
\begin{center}
\leavevmode
\hbox{
\epsfysize=6.0in
\epsffile{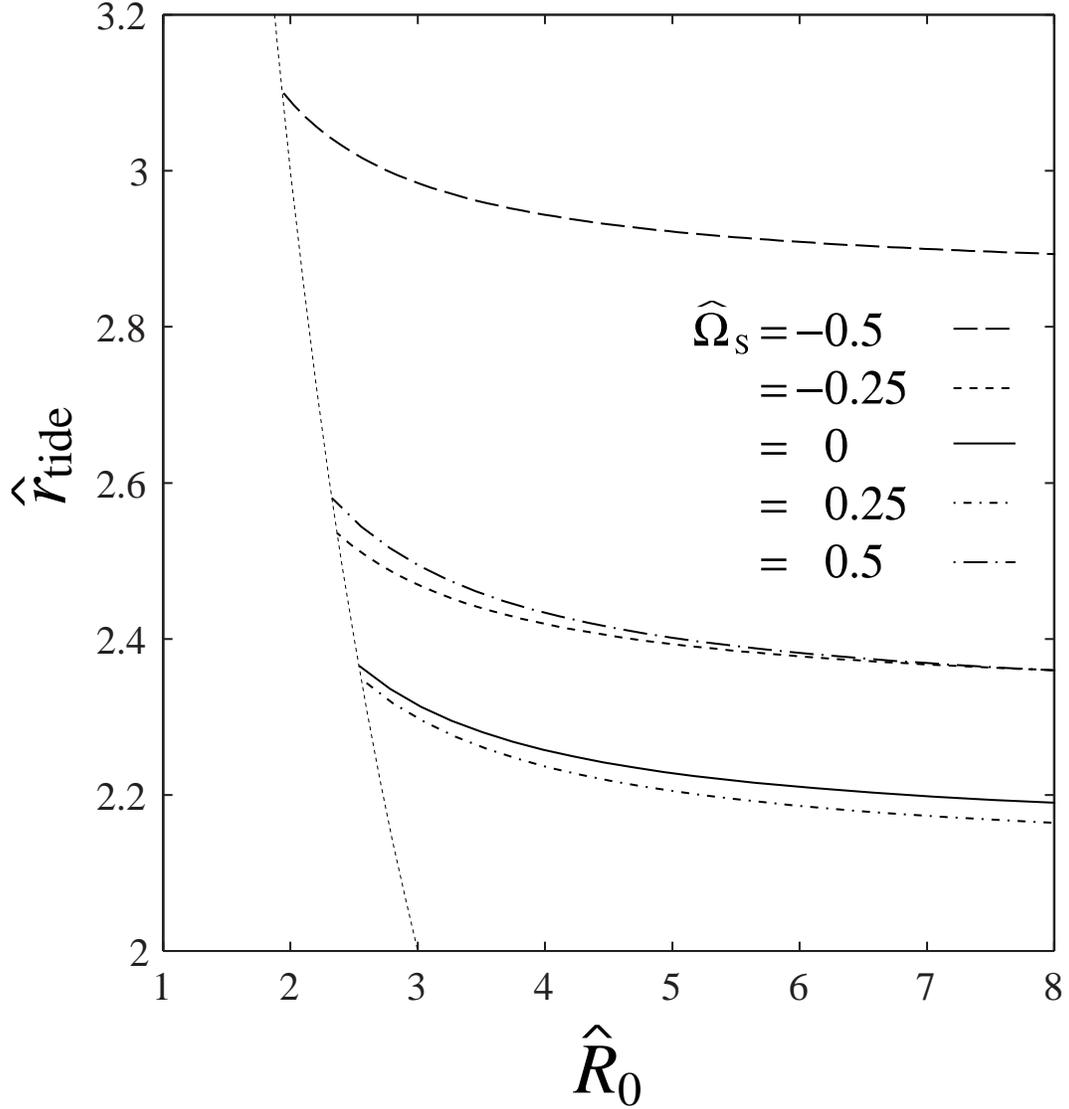}}
\caption{
Tidal limits in models with initial rotation ($\hat \Omega_s\ne 0$) 
for $n=0.5$ and $\hat a=0$.  As in Figs.~4-5, each curve
is truncated at $\hat R_0=\hat R_{0,{\rm crit}}$, where
$r_{\rm tide}=r_{\rm isco}$ [The short-dashed line represents the 
ISCO, i.e., $\hat r\hat R_0=(r/M)_{\rm isco}$].  
\label{fig:omega}}
\end{center}
\end{figure}

To illustrate the dependence of the tidal limit on the initial stellar
spin, Figure \ref{fig:omega} shows $\hat r_{\rm tide}$ as a function $\hat R_0$
for $n=0.5$, $\hat a=0$ and several different values of $\hat\Omega_s$. For
negative $\hat\Omega_s$ (positive ${\hat{\cal C}}$), the internal rotation
$\Lambda$ is larger (compared to the ${\hat{\cal C}}=0$ case), and this
induces larger distortion to the star, and thus the star gets disrupted
at a larger radius. For small, positive $\hat \Omega_s$, the internal rotation
$\Lambda$ is reduced, and thus $\hat r_{\rm tide}$ is smaller. 
But as $\hat \Omega_s$ becomes larger ($\hat\Omega_s\go 0.3$), 
$\hat\Lambda$ also gets larger and $\hat r_{\rm tide}$ is increased over the
$\hat \Omega_s=0$ value. 

\subsection{Roche Model: Corotating Binaries}

Although it is likely that equilibrium sequences with constant circulation 
should describe the quasi-static evolutionary phase of NS-BH binaries (see the
beginning of \S 4 for discussion), some degree of spin-orbit synchronization 
can still be achieved at small orbital radii through an anomalously high
(turbulent) viscosity (so that $\sin 2\alpha$ is not much less than unity).
This is especially true for WD-BH binaries, for which
corotating models (with $\Lambda=0$) may be relevant. 
For completeness, Table 2 lists selected values of the tidal 
limit (of ``Roche limit'') as a function of $\hat R_0$ and $\hat a$
for polytropic index $n=0$ and $n=1.5$. For the incompressible model
($n=0$), our result agrees with that of Fishbone (1973) (although
a different parametrization was adopted in Fishbone's paper). 
Similar to the irrotational models, we find that the GR effect makes
$\hat r_{\rm tide}$ larger than the Newtonian value,  
and that $\hat r_{\rm tide}$ is larger for $\hat a=-1$ than for $\hat a=1$.

\section{Tidal Effect on the Orbit}

The previous sections are based on the assumption that the star $m$ moves
in a geodesic of the BH metric. In reality, there are two 
corrections to the geodesic. 
First, even if $m$ were a point mass, its finite mass can modify the 
spacetime of the system. For $r\gg M+m$, the problem can be solved by
post-Newtonian expansion (e.g., Blanchet et al.~1998 and 
references therein) 
or BH perturbation theory (e.g., Mino et al.~1997); when $r$ is comparable to
$M+m$, nonperturbative techniques are needed (see Kidder, Will 
\& Wiseman 1993, Buonanno \& Damour 1998 and references therein) 
and definitive result will only come from numerical
relativity. Second, there is a correction due to the finite size of the star.
Here we focus on the second correction, i.e., the tidal 
effect on the orbit.

The tidal interaction energy is given by 
\be
W_{\rm tide}=-{M\over 2r^3}\left(2I_{11}-I_{22}-I_{33}\right)
-{3M\over 2r^3}\left({K\over r^2}\right)\left(I_{11}-I_{33}\right),
\ee
(see eq.~[\ref{wtide}]). The dimensionless ratio
\be
\delta={W_{\rm tide}\over (-Mm/r)}
={\kappa_n\over 10}\left({m\over M}\right)^{2/3}{1\over\hat r^2}
\left[\left(2\hat a_1^2-\hat a_2^2-\hat a_3^2\right)
+3\left({K\over r^2}\right)\left(\hat a_1^2-\hat a_3^2\right)\right],
\ee
measures the correction of the tidal effect on the orbital motion
(deviation from geodesic)\footnote{In Newtonian theory, the orbital
angular freqency is given by $\Omega=[(M+m)/r^3]^{1/2}(1+3\delta/2)$.}.
Since $(2\hat a_1^2-\hat a_2^2-\hat a_3^2)$ and
$(\hat a_1^2-\hat a_3^2)$ are of order $1/\hat r^3$, and $K/r^2$ 
is of order $M/r<1$, we see that $\delta$ scales as $1/\hat r^5$. 
Note that even near the tidal radius ($\hat r\sim 1$), $\delta$ can be
quite small if $m/M$ is small. 

In the absence of tidal coupling, the dynamical instability limit 
$r_{\rm dyn}$ of the orbit is located at ISCO. We assume that
$r_{\rm isco}$ is given by the standard $M\gg m$ result.
(When $m/M$ is finite, see Buonanno \& Damour 1999 and references therein.)
Tidal coupling is expected to shift $r_{\rm dyn}$ outward compared to
$r_{\rm ISCO}$ (see Lai \& Wiseman 1996). To estimate the correction,
$(r_{\rm dyn}-r_{\rm isco})$, we write the total equilibrium energy of
the binary system as
\be
{\cal E}=mE+\Delta {\cal E},
\ee
where $E$ is the energy per unit mass of a test particle orbiting a
Kerr BH (see eq.~[\ref{equation:energy}]), and $\Delta {\cal E}$
is the correction due to the finite size of the star. Minimizing
${\cal E}$ would give $r_{\rm dyn}$. Unfortunately there is no unambiguous
way of writing down the total energy ${\cal E}$ in general relativity.
However, we can still attempt to determine how 
$(r_{\rm dyn}-r_{\rm isco})$ scales with various parameters of the system. 
From Newtonian consideration, $\Delta {\cal E}$ is certainly 
of order $(-W_{\rm tide})$ --- The negative sign arises because
the tidal attraction increases the kinetic energy of the orbital motion, 
which more than compensates for the neative tidal potential energy.
We therefore set $\Delta{\cal E}=-\beta W_{\rm tide}$, with $\beta$ of order
unity. Assuming that $r_{\rm dyn}$ is close to $r_{\rm isco}$, 
we find\footnote{In taking derivative of $W_{\rm tide}$ with respect to
$r$, we have assumed that $K/r^2$ is roughly constant, which we have
checked to be reasonable as long as $\hat a$ is not too close to 1. 
Also we have found that $(2 \hat a_1^2-\hat a_2^2-\hat a_3^2)$ is 
approximately equal to twice of $(a_1^2-a_3^2)$, and both scale 
approximately as $r^{-3}$. Relaxing these approximations would give rise to a
factor of order unity which we can absorb into the definition of $\beta$.}:
\be
\epsilon \equiv {r_{\rm dyn}-r_{\rm isco}\over M}
=C \beta \kappa_n 
\hat R_0^2 \left({m \over M}\right)^{2/3}
\left(2\hat a_1^2-\hat a_2^2-\hat a_3^2\right)_{\rm isco},
\label{dyn}\ee 
where $(2\hat a_1^2-\hat a_2^2-\hat a_3^2)$ is evaluated at $r=r_{\rm isco}$.
The constant $C$ depends on $r_{\rm isco}/M$ and $d^2E/dr^2$ (evaluated at
$r_{\rm isco}$):  $C=0.116$ for $\hat a=0.998$, and $0.106$ for 
$\hat a=0$ and $0.077$ for $\hat a=-1$. 
Although we cannot determine the precise value of $\beta$,
we expect the scaling in eq.~(\ref{dyn}) to be valid in general.

The biggest correction to $r_{\rm dyn}$ occurs when tidal disruption occurs 
just inside the ISCO. For example, when $\hat R_0=2$, $a=0$ and $n=0$,
we have $\epsilon = 0.53 \beta (m/M)^{2/3}$ (assuming $m/M$ is small).

\section{Possible Applications to Gamma-Ray Bursts}

In this section we discuss possible applications of our results 
to Gamma-Ray Bursts (GRBs).

\subsection{Black Hole -- Neutron Star Binaries}

Coalescence of BH-NS binaries has frequently been invoked as
the central engine of GRBs (e.g., Narayan, Paczynski \&
Piran 1992; Meszaros, Rees \& Wijers 1998; Piran 1998). It is expected 
that a thick disk or torus will result from the disruption 
of the NS. To account for the temporal properties of observed GRBs, 
the disk must be long-lived (compared to the orbital timescale)
(e.g., Meszaros, Rees \& Wijers 1998; van Putten 1999), and therefore
the tidal disruption must occur outside the ISCO, i.e., $r_{\rm tide}>r_{\rm
ISCO}$.  Approximating the NS as a $n=0.5$ polytrope, we find that the tidal
radius for critical $\hat R_0=\hat R_{0,{\rm crit}}$
(where $r_{\rm tide}=r_{\rm isco}$) is $\hat r_{\rm tide}=2.56$ 
(see Table 1). Therefore the condition $r_{\rm tide}>r_{\rm isco}$ translates
to 
\be
{M\over m}<\left({M\over m}\right)_{\rm crit}=
3.12\left({R_0\over 5m}\right)^{3/2}\!\left({6M\over r_{\rm
isco}}\right)^{3/2}.
\ee
For canonical neutron star parameters ($m=1.4M_\odot,~R_0=10$~km, 
corresponding to $R_0/m\simeq 4.84$), tidal disruption occurs outside 
ISCO only if $M\lo 4.4M_\odot$ (for $\hat a=0$ BH) or 
$M\lo 64M_\odot$ (for $\hat a=1$ BH). 

\begin{figure}[hbt]
\begin{center}
\leavevmode
\hbox{
\epsfysize=6.0in
\epsffile{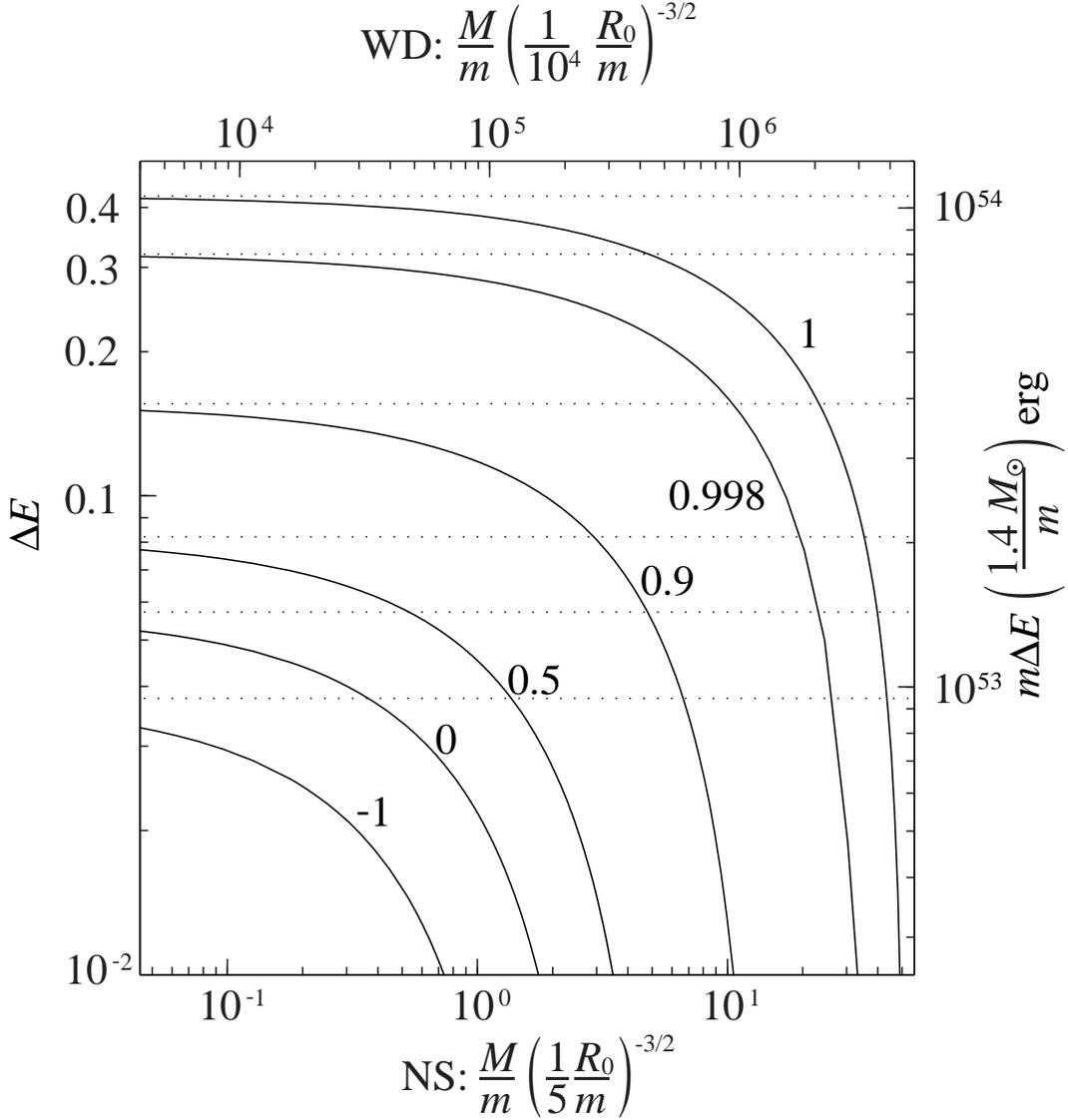}}
\caption{
The maximum energy available (per unit mass of accreting matter),
$\Delta E=E(r_{\rm tide})-E(r_{\rm isco})$, 
in the accretion disk following the tidal disruption, is plotted
against the ``effective'' mass ratio.
The numerical value adjacent to each solid curve indicates the
Kerr parameter $\hat a$. ($\hat a$ = 0.998 is the Thorne limit for 
Kerr BHs fed by accretion disks.)  
The dotted horizontal lines indicate the limiting energy for 
disruption at $\infty$ for each of the Kerr parameters.  
For BH-NS binaries, the bottom horizontal axis shows the mass ratio scaled
for $R_0/m=5$ (Results for $M/m$ less than a few can only be viewed 
suggestively, and results below $M/m=1$ is not physically meaningful).  
For BH-WD binaries, the top horizontal axis shows the mass ratio scaled for
$R_0/m=10^4$. The label on the right gives the total energy in ergs
for $m=1.4M_\odot$. Note that in this figure, $r_{\rm tide}$ is calculated
for $n=0$, but the dependence of $\Delta E$ on $n$ is weak.
\label{fig:mn}}
\end{center}
\end{figure}

We can calculate the maximum energy available in the disk following
a disruption by comparing the binding energies at $r_{\rm
tide}$ and at $r_{\rm isco}$. Assuming that the disk material 
follows geodesic, the maximum energy release 
per unit mass of accreting gas is given by
\be
\Delta E = E(r_{\rm tide})-E(r_{\rm isco}),    
\ee
where $E(r)$ is given by eq.~(\ref{equation:energy}). Note that $\Delta E$
depends only on $\hat R_0$. Figure \ref{fig:mn} depicts $\Delta E$ as a
function of the ``effective'' mass ratio
$(M/m)(R_0/5m)^{-3/2}=(5/\hat R_0)^{3/2}$
for different value of BH spin parameter $\hat a$
($\hat a=-1,\,0,\,0.5,\,0.9,\,0.998$ and $a\rightarrow 1$;
Note that $\hat a=0.998$ corresponds to the maximum value of $\hat a$ 
that can be achieved in accretion; Thorne 1974). 
Clearly, $\Delta E=0$ when $M/m=(M/m)_{\rm crit}$ since $r_{\rm
tide}=r_{\rm isco}$. When $(M/m)\ll (M/m)_{\rm crit}$, $\Delta E$ approaches
the standard radiation efficiency for thin disk, $\Delta E=1-E(r_{\rm isco})$.
The dependence of $\Delta E$ on $\hat a$ mainly comes from 
$r_{\rm isco}$ since $r_{\rm tide}$ is relatively insensitive to
$\hat a$ (see Figs.~\ref{tidal_1}-\ref{tidal_2} and Table 1). 
Note that our calculation is valid only when $M/m\gg 1$,
so the results for $M/m$ less than a few should be considered suggestive 
only. 

Recent observations indicate that some GRBs are highly energetic,
with the apparent emitted energy in gamma rays reaching 
$10^{53}$ erg (e.g., Kulkarni et al.~1999 and references therein). 
Uncertainty in beaming precludes accurate
determination of the true gamma ray energy. It has been 
suggested that the total energy in the fireball may be as 
high as $10^{54}$ erg (e.g., Kumar 1999). If this number holds up,
we may rule out NS-BH binary coalescence as the central engine
of GRBs (see Fig. ~\ref{fig:mn}).

\subsection{Black Hole -- White Dwarf Binaries}

For a BH-WD binary, tidal disruption occurs outside ISCO ($r_{\rm tide}>
r_{\rm isco}$) when
\be
{M\over m}<\left({M\over m}\right)_{\rm crit}=
2.18\times 10^5\left({R_0\over 10^4m}\right)^{3/2}\!\left({6M\over 
r_{\rm isco}}\right)^{3/2},
\ee
where we have adopted the $n=1.5$ polytropic model for the WD, and
used $\hat r_{\rm tide}=2.173$ (see Table 1) for 
$\hat R_0=\hat R_{0,{\rm crit}}$.
Although there is no evidence for massive BHs with mass
around $10^5M_\odot$ (e.g., Richstone et al.~1998), the observational
constraint on such BHs is also weak. If they do exist, 
they may capture WDs and the merger of the BH-WD binary may lead to 
a GRB (R. Blandford 1997, private communication). 
For $(M/m)\ll (M/m)_{\rm crit}$, a Newtonian description for tidal 
disruption is adequate; but when $M/m$ is close to $(M/m)_{\rm crit}$, 
a general relativistic treatment is necessary. Since $M\gg m$ is well
satisfied, the model presented in this paper should provide an accurate
description of BH-WD binaries. 

Figure \ref{fig:mn} shows the maximum energy available in a BH-WD 
tidal disruption. Unlike BH-NS binaries, the limiting value of 
$\Delta E$, corresponding to the thin disk limit, 
$\Delta E=1-E(r_{\rm isco})$, can be easily achieved for a large range of 
mass ratios. 

\section{Discussion}

Quasi-equilibrium models of neutron star binaries provide
an accurate description of the late stage of inspiralling 
binary stars prior to the dynamical merger. 
While quasi-equilibrium NS-NS binary models (the so-called ``Darwin-Riemann
problems'') have been successfully constructed in general relativity 
(see Bonazzola et al.~1999a,b for a review), only Newtonian models 
of NS-BH binaries have been studied so far (see Uryu \& Eriguchi 1999).
The semi-analytic Roche-Riemann models studied in this paper 
provide useful insight on the relativistic tidal effects in coalescing
NS-BH and WD-BH binaries. 

We note that our model is exact when the following conditions are satisfied:
(i) $M\gg m$ [so that the spacetime is given by the Kerr metric of the BH
(\S 2), and the tidal correction to the orbit is negligible (\S 5)];
(ii) $R_0/m\gg 1$ (so that we can treat the fluid star as a Newtonian object);
(iii) The radius of curvature $(r^3/M)^{1/2}\gg R_0$ [so that high-order 
tidal effect can be neglected (\S 2); At the $r=r_{\rm tide}\sim
(M/m)^{1/3}R_0$, this condition translates to $(R_0/m)^{1/2}\gg 1$, and
is equivalent to condition (ii)];
(iv) $n=0$ [The ellipsoid ansatz is exact when the external force is a linear
function of spatial coordinates, as required by (iii), and when 
the fluid is incompressible (Chandrasekhar 1969)].
Clearly, for WD-BH binaries, conditions (i)-(iii) are well-satisfied, 
and the ellipsoid ansatz is a good approximation for polytrope of 
$n=1.5$ (appropriate for low-mass WD) (LRS1\footnote{Note that
for very compressible fluid (large $n$), the tidal limit corresponds to
the point where mass transfer starts; see Uryu \& Eriguchi (1999).}).
Therefore our findings that general relativitic effect increases
$r_{\rm tide}$ and that $r_{\rm tide}$ is largest for $\hat a=-1$ (regtrograde
BH) are expected to hold. 

For NS-BH binaries, conditions (i)-(iii) are satisfied only
approximately, we must apply our results with caution. 
In particular, post-Newtonian correction to the self-gravity is 
likely to be important. The height $H$ of tidal bulge on the star 
can be estimated from 
$(MR_0/r^3)\left[1+{\cal O}(K/r^2)\right]\sim
(mH/R_0^3)\left[1+{\cal O}(m/R_0)\right]$. Setting $H\sim R$ gives
the tidal radius
\be
r_{\rm tide} \sim R_0\left( {M \over m } \right)^{1/3}
\left[ 1 + {\cal O} \left({K\over r^2}\right) -{\cal O}\left({m\over
R_0}\right)\right].
\ee    
While relativistic tidal field (the $K/r^2$ term) increases $r_{\rm tide}$,
the post-Newtonian self-gravity effect (the $m/R_0$ term) tends to
reduce $r_{\rm tide}$. Since $m/R_0$ for a NS is of order $20\%$, 
comparable to $K/r^2$ (which is always less than $1/3$; see Fig.~3),
the post-Newtonian self-gravity effect cannot be neglected. 
Note that our finding that $r_{\rm tide}$ is largest for 
$\hat a=-1$ is still expected to be valid.

\acknowledgments
P.W. acknowledges support from the NSF REU program at Cornell, 
and D.L. thanks the Alfred P. Sloan Foundation for a research fellowship.


\clearpage

\begin{deluxetable}{rrrr}
\tablecolumns{4}
\tablewidth{0pc}
\tablecaption{Tidal Limits for Irrotational Binaries}
\tablehead{
\colhead{}    &  \multicolumn{3}{c}{$\hat r_{\rm tide}$} \\
\cline{2-4} \\
\colhead{$\hat R_0$} & \colhead{$\hat a=-1$}   & \colhead{$\hat a=0$}    &
\colhead{$\hat a=+1$}
} 
\startdata
\cutinhead{$n=0$}
$\infty$ & 2.502   	& 2.502   	& 2.502 \\*
6        & 2.607   	& 2.569  	& 2.539 \\*
4        & 2.689   	& 2.611   	& 2.552 \\*
3.265    & 2.757   	& 2.644   	& 2.561 \\* 
3        & \nodata 	& 2.661   	& 2.565 \\*
2.176    & \nodata 	& 2.757   	& 2.581 \\* 
2        & \nodata 	& \nodata 	& 2.586 \\* 
1        & \nodata 	& \nodata 	& 2.638 \\* 
0.363    & \nodata 	& \nodata 	& 2.757 \\
\cutinhead{$n=0.5$}
$\infty$ & 2.320 	& 2.320 	& 2.320 \\*
6        & 2.430 	& 2.389 	& 2.357 \\*
4        & 2.517 	& 2.433 	& 2.370 \\*
3.515    & 2.560 	& 2.454 	& 2.375 \\* 
3        & \nodata      & 2.487 	& 2.382 \\*
2.344    & \nodata      & 2.560 	& 2.395 \\* 
2        & \nodata      & \nodata      	& 2.404 \\* 
1        & \nodata      & \nodata      	& 2.455 \\* 
0.390    & \nodata      & \nodata      	& 2.560 \\
\cutinhead{$n=1$}
$\infty$ & 2.140 	& 2.140 	& 2.140 \\*
6        & 2.255 	& 2.210 	& 2.176 \\*
4        & 2.349 	& 2.257 	& 2.190 \\*
3.805    & 2.366 	& 2.266 	& 2.191 \\* 
3        & \nodata      & 2.315 	& 2.202 \\*
2.536    & \nodata      & 2.366 	& 2.210 \\* 
2        & \nodata      & \nodata      	& 2.223 \\* 
1        & \nodata      & \nodata      	& 2.273 \\* 
0.423    & \nodata      & \nodata      	& 2.366 \\
\cutinhead{$n=1.5$}
$\infty$ & 1.960 	& 1.960 	& 1.960 \\*
6        & 2.083 	& 2.034 	& 1.997 \\*
4.141    & 2.173 	& 2.079 	& 2.009 \\*
4        & \nodata      & 2.085 	& 2.010 \\* 
3        & \nodata      & 2.148 	& 2.023 \\*
2.761    & \nodata      & 2.173 	& 2.026 \\* 
2        & \nodata      & \nodata      	& 2.044 \\* 
1        & \nodata      & \nodata      	& 2.093 \\* 
0.460    & \nodata      & \nodata      	& 2.173 \\
\enddata
\tablecomments{
$\hat{R}_0\equiv {R_0\over m}\left({m\over M}\right)^{2/3}$
($m,~R_0$ are the mass and radius of the star, $M$ is the BH mass),
$\hat r_{\rm tide}\equiv {r_{\rm tide} \over R_0}
\left({m\over M}\right)^{1/3}$, $\hat a$ is the Kerr parameter of the BH,
and $n$ is the polytropic index. 
The last item in each column corresponds to 
$\hat R_0=\hat R_{0,{\rm crit}}$ at which $r_{\rm tide}=r_{\rm isco}$ or
$\hat{R}_0 \hat r_{\rm tide}=\left({r\over M}\right)_{\rm isco}$.}
\end{deluxetable}

\clearpage

\begin{deluxetable}{rrrr}
\tablecolumns{4}
\tablewidth{0pc}
\tablecaption{Tidal Limits for Corotating Binaries}
\tablehead{
\colhead{}    &  \multicolumn{3}{c}{$\hat r_{\rm tide}$} \\
\cline{2-4} \\
\colhead{$\hat R_0$} & \colhead{$\hat a=-1$}   & \colhead{$\hat a=0$}    &
\colhead{$\hat a=+1$} \\
} 
\startdata
\cutinhead{$n=0$}
$\infty$ & 2.455   	& 2.455   	& 2.455 \\*
6        & 2.566   	& 2.525  	& 2.493 \\*
4        & 2.653   	& 2.570   	& 2.508 \\*
3.311    & 2.718  	& 2.602   	& 2.516 \\* 
3        & \nodata 	& 2.623   	& 2.521 \\*
2.208    & \nodata 	& 2.718   	& 2.537 \\* 
2        & \nodata 	& \nodata 	& 2.543 \\* 
1        & \nodata 	& \nodata 	& 2.597 \\* 
0.368    & \nodata 	& \nodata 	& 2.718 \\
\cutinhead{$n=1.5$}
$\infty$ & 2.035 	& 2.035 	& 2.035 \\*
6        & 2.148 	& 2.103 	& 2.069 \\*
4.018    & 2.240 	& 2.149 	& 2.082 \\*
4        & \nodata      & 2.150 	& 2.082 \\* 
3        & \nodata      & 2.207 	& 2.093 \\*
2.679    & \nodata      & 2.240 	& 2.099 \\* 
2        & \nodata      & \nodata      	& 2.113 \\* 
1        & \nodata      & \nodata      	& 2.160 \\* 
0.447    & \nodata      & \nodata      	& 2.240 \\

\enddata
\tablecomments{
$\hat{R}_0\equiv{R_0\over m}\left({m\over M}\right)^{2/3}$, 
$\hat r_{\rm tide}\equiv {r_{\rm tide} \over R_0}
\left({m\over M}\right)^{1/3}$. The last item in each column corresponds to 
$\hat{R}_0=\hat R_{0,{\rm crit}}$ at which $r_{\rm tide}=r_{\rm isco}$ or 
$\hat{R}_0 \hat r_{\rm tide}=\left({r\over M}\right)_{\rm isco}$.}
\end{deluxetable}

\clearpage

\end{document}